\documentclass[12pt]{article}
\usepackage{amsmath, amssymb, amsfonts, amsthm}
\usepackage[margin=1in]{geometry}
\usepackage{graphicx}
\usepackage{subcaption}
\usepackage{float} 
\usepackage{array}
\usepackage[latin1]{inputenc}
\usepackage{enumerate}
\usepackage{xspace}
\usepackage{caption}
\usepackage{booktabs}
\usepackage{color}
\usepackage{lmodern}
\usepackage{tikz}
\usepackage{comment}
\usepackage{enumitem}
\usepackage{pgfplots}
\usepackage{lipsum}
\usepackage{algorithm}
\usepackage{algpseudocode}
\usepackage{natbib}
\usepackage{xcolor}
\usepackage{color-edits}
\graphicspath{ {./images/} }
\usepackage{amsthm}
\usetikzlibrary{shapes, snakes, backgrounds, calc}

\addauthor{de}{red}     
\addauthor{mg}{blue}    

\setlist[itemize]{
    topsep=0pt,     
    partopsep=0pt,  
    itemsep=0pt,    
    parsep=0pt,     
    leftmargin=*    
}
\usepackage{authblk}
\title{An Agglomerative Clustering Algorithm for Simulation Output Distributions Using Regularized Wasserstein Distance}

\author[1]{\small Mohammadmahdi Ghasemloo}
\author[1]{\small David J. Eckman}

\affil[1]{\small Department of Industrial and Systems Engineering, Texas A\&M University, College Station, TX 77843, United States.}
\affil[ ]{\small \texttt{mohammad\_ghasemloo@tamu.edu}, \texttt{eckman@tamu.edu}}

\makeatletter
\renewcommand{\@maketitle}{
  
  \vskip -30em 
  \begin{center}%
    \let \footnote \thanks
    {\LARGE \@title \par}%
    \vskip 1em 
    {\large
      \lineskip .1em%
      \begin{tabular}[t]{c}%
        \@author
      \end{tabular}\par}%
    \vskip -5em 
  \end{center}%
  \par
  \vskip 1em}
\makeatother
\usepackage{titlesec}

\titlespacing{\section}{0pt}{1.5ex plus 1ex minus .2ex}{0.5ex plus .2ex}
\titlespacing{\subsection}{0pt}{1.0ex plus 0.5ex minus .2ex}{0.5ex plus .2ex}
\begin{document}
\maketitle

\begin{abstract}%
Using statistical learning methods to analyze stochastic simulation outputs can significantly enhance decision-making by uncovering relationships between different simulated systems and between a system's inputs and outputs. We focus on clustering multivariate empirical distributions of simulation outputs to identify patterns and trade-offs among performance measures. We present a novel agglomerative clustering algorithm that utilizes the regularized Wasserstein distance to cluster these multivariate empirical distributions. This framework has several important use cases, including anomaly detection, pre-optimization, and online monitoring. In numerical experiments involving a call-center model, we demonstrate how this methodology can identify staffing plans that yield similar performance outcomes and inform policies for intervening when queue lengths signal potentially worsening system performance.
\end{abstract}%

\section{Introduction}

A simulation model is a computational representation of a real-world system designed to study its behavior under various settings or scenarios. Simulation models are used extensively in fields such as engineering, economics, healthcare, and environmental science to predict and analyze outcomes of different scenarios without the need to experiment in the real world, which can be costly, time-consuming, or impractical. 
Outputs of a simulation model typically correspond to key performance indicators (KPIs) of interest to the decision maker, e.g., profit, throughput, or service level.
For stochastic simulation models, simulating a given scenario generates outputs that vary from replication to replication, thus each scenario has an associated probability distribution describing the stochastic behavior of its outputs.
This distribution can be approximated by the empirical distribution produced by the output data.
When conducting simulation experiments, the user decides which scenarios are simulated and how many replications are run. 
Through the control of pseudo-random numbers, simulation experiments can be easily designed to produce output data that satisfies the standard assumptions of being independent and identically distributed, in contrast to most data obtained from the real world.
Common tools for analyzing simulation output data include basic summary statistics (e.g., sample means, variance, and covariances) and visualization tools (e.g., histograms and boxplots). For problems with multiple KPIs, the multivariate empirical distribution produced by the data contains valuable information about system performance, but can be difficult to analyze and plot.
To reveal important patterns and relationships that cannot be detected by conventional data analysis methods, we propose clustering the empirical distributions of simulated scenarios.

Clustering is an unsupervised learning approach that can help discover important patterns and relationships in complex datasets \citep{james2013introduction}.
In the context of simulation output data, clustering can identify scenarios with similar KPIs, or more precisely, similar output distributions. 
By understanding the characteristics of each cluster of distributions, a decision maker can draw meaningful comparisons of clusters and make informed decisions. 
We consider three important applications of clustering for enhancing simulation output analysis: {\it anomaly detection}, which involves identifying and investigating outliers in simulation outputs; {\it pre-optimization}, which involves formulating simulation-optimization problems and identifying promising initial solutions; and {\it online monitoring}, which involves tracking the system state over time and using classification methods to detect potentially undesirable system behavior and trigger appropriate actions or alerts. Our approach enhances decision making by extracting insights from complex simulation data with broad industry applications. For example, in call center optimization, our clustering algorithm can quickly distinguish staffing configurations that improve service levels and reduce costs. In healthcare operations, clustering can identify efficient resource allocation strategies that improve patient care and aid in evaluating trade-offs among metrics like wait times, resource utilization, and costs. In portfolio management, clustering can help with risk assessment and strategic planning by highlighting patterns in investment returns and enabling targeted interventions to prevent loss across different market conditions.





We propose an agglomerative clustering method that uses the complete-linkage criterion for forming clusters.
We choose agglomerative clustering for its flexibility, as it does not require specifying a predetermined number of clusters, and opt for complete linkage because it maintains compact and well-separated clusters. 
For the algorithm's measure of dissimilarity between distributions, we choose the regularized Wasserstein distance. 
The (unregularized) Wasserstein distance quantifies the distance between two discrete probability distributions by the minimum amount of ``work''---defined as the product of the probability mass that needs to be moved and the distance it needs to be transported---required to transform one distribution into the other \citep{villani2009optimal}.
The Wasserstein distance is chosen over other metrics like Kullback-Leibler divergence or Jensen-Shannon divergence due to its ability to handle distributions with non-overlapping supports, as arises when working with continuous-valued empirical distributions.
Another advantage of the Wasserstein distance is the notion of a barycenter, which acts like an average among distributions. Barycenters are particularly useful for post-hoc analysis, when desiring to summarize each cluster with a representative distribution \citep{peyre2019computational, arjovsky2017wasserstein, cuturi2013sinkhorn}.
However, computing the Wasserstein distance between two discrete distributions entails solving a linear program \citep{villani2009optimal}, which could become computationally intensive when working with large datasets. 
The regularized Wasserstein distance, also known as the Sinkhorn distance, adds an entropic regularization term that promotes smoother transport plans, is easier to compute, and more stable, making it well suited for data-intensive applications \citep{cuturi2013sinkhorn}. 
Our algorithm determines the optimal number of clusters based on the silhouette index \citep{shahapure2020cluster}, a centroid-free metric that evaluates the quality of a clustering in terms of its intra-cluster and inter-cluster distances.


Clustering has been used for many purposes in many fields, such as analyzing gene expression in bioinformatics \citep{eisen1998cluster}, market segmentation in economics \citep{wedel2000market}, and textures and shapes in image processing \citep{pappas1989adaptive}. 
Hierarchical clustering techniques, specifically agglomerative clustering, build nested clusterings by iteratively merging the two most similar clusters \citep{murtagh2012algorithms} and produce a dendrogram that depicts the nested clusterings at different levels \citep{jain1999data}. Adaptations like single-linkage, complete-linkage, and average-linkage offer great flexibility in how agglomerative clustering algorithms merge clusters \citep{jain2010data}. More recently, clustering has been employed in the study of stochastic simulation models, including for simulation optimization \citep{li2024efficient,zhang2024sample,peng2018efficient} and reducing model-form uncertainty \citep{abdallah2020unsupervised}. To the best of our knowledge, we are the first to propose agglomerative clustering of simulation output distributions and investigate important use cases.

The Wasserstein distance is used extensively in machine learning applications, such as in computer vision and pattern recognition to robustly compare visual feature distributions \citep{rubner2000earth}. For large-scale problems, variations like the regularized Wasserstein distance improve computational efficiency. For example, \cite{benamou2015iterative} employ iterative Bregman projections to efficiently solve regularized transportation problems, thereby improving scalability and accuracy. Our clustering framework uses the algorithm of \cite{benamou2015iterative} to calculate the regularized Wasserstein distance and regularized Wasserstein barycenters. For a more detailed overview of computational techniques for optimal transport and its practical applications, we direct the reader to \cite{peyre2019computational}. 

The Wasserstein distance has been applied in $k$-means clustering, particularly in the identification of market regimes \citep{horvath2021clustering} and the analysis of financial data \citep{riess2023geometry}. Instead of relying on standard $k$-means, \cite{henderson2015ep} proposed an extension to cluster one-dimensional, continuous-valued empirical distributions. However, as sample sizes grow, their EP-MEANS algorithm becomes less efficient and struggles to generalize to multivariate cases. A different approach was introduced by \cite{irpino2014dynamic}, who developed adaptive algorithms using squared Wasserstein distances, leading to improved performance in both synthetic and real-world scenarios. Meanwhile, \cite{del2019robust} focused on enhancing robustness and efficiency by applying trimmed $k$-means on distributions with the squared Wasserstein distance. \cite{zhuang2022wasserstein} expanded $k$-means to the Wasserstein space, though they encountered issues with irregularity and non-robustness in barycenter-based clustering methods. Our hierarchical clustering algorithm addresses many of these limitations, particularly in terms of regularity and scalability. Another hierarchical clustering method based on optimal transport distance measures is offered by \cite{chakraborty2020hierarchical}, though their work does not directly target the clustering of distributions as ours does.



The rest of this paper is organized as follows: In Section~\ref{sec:usecases}, we elaborate on use cases of clustering simulation output distributions. In Section~\ref{sec:clustering}, we introduce the relevant notation and propose our agglomerative clustering algorithm. We present the results of several experiments on a call-center staffing problem in Section~\ref{sec:experiments} and conclude in Section~\ref{sec:conclusion}.

\section{Use Cases}
\label{sec:usecases}

Before presenting the algorithm itself, we first discuss the utility and versatility of clustering simulation output distributions through several use cases.

\paragraph{Anomaly Detection} In the context of simulation experiments, anomalies can be categorized as artificial or systemic. An artificial anomaly is typically associated with logic or coding errors within the simulation model, whereas a systemic anomaly is related to inherent features of the system. When using hierarchical clustering algorithms, anomalous output distributions can be identified by examining the dendrogram, the distances between clusters, or the cluster sizes \citep{loureiro2004outlier}. After identifying an anomalous output distribution, the decision maker would first scrutinize the simulation code to determine if it is an artificial anomaly. If the anomaly is not artificial, then he or she might investigate further by, for example, examining the marginal distributions, correlation matrices, and corresponding input variables. Hierarchical clustering algorithms such as ours are expected to be more stable for identifying outliers than non-hierarchical clustering methods whose clusterings strongly depend on the initialization of the clusters.

\paragraph{Pre-Optimization} In many practical situations, there are tradeoffs between KPIs, and the decision maker may be unable to articulate a priori what constitutes ``good'' versus ``bad'' system performance. By clustering output distributions and obtaining the barycenters, the decision maker can be presented with a more manageable number of distributions to compare. The decision maker could then conduct a series of pairwise comparisons of barycenters and thereby eliminate some clusters based on unformalized notions of preferred performance until a small number of clusters (or scenarios) remain. The clustering analysis can also help the decision maker specify which metrics should be modeled as objectives in a subsequent simulation-optimization problem and which should be treated as constraints. Achievable thresholds for the constraints can be set based on the observed performance outcomes of the simulated scenarios. Additionally, by examining the inputs associated with scenarios in a promising cluster, the decision maker can identify promising regions of the input space from which to initiate an optimization search, potentially leading to more rapid progress toward the optimal solution.

\paragraph{Online Monitoring} This application concerns how the output of a simulation model is influenced by state variables, namely, those that evolve over time and can be observed, but not directly controlled, by the decision maker. We envisage an online monitoring framework in which clustering is performed offline and state variables are later tracked in real time and classification algorithms are utilized to help the decision maker anticipate changes in system performance. This approach involves a preliminary simulation experiment in which the scenarios correspond to different initial states, followed by the clustering of the generated outputs.
When monitoring the system's state online, classification algorithms can be used to predict the cluster to which an observed state's output distribution may belong. Conversely, if the classification algorithm struggles to assign a state to a single cluster, such as a tie when using $k$-nearest neighbors, it could indicate that the system's performance may change soon, potentially prompting intervention.

\section{Clustering Simulation Output Distributions}
\label{sec:clustering}

Suppose there are $N$ scenarios under consideration, and for each Scenario $i$, $i=1,2,\ldots,N$, we obtain $n_i$ independent simulation replications. We make no assumptions about whether the replications performed at different systems are independent. In the numerical experiments, we examine the use of common random numbers (CRN) to induce positive correlation in the outputs from different systems and observe its effect on our clustering technique. Let $ \mathbf{y}_{il} \in \mathbb{R}^d$ denote the vector output of the $ l $th simulation replication at Scenario $ i $ and let $\mu_i := n_i^{-1} \sum_{l=1}^{n_i}  \delta_{\mathbf{y}_{il}}$ denote the corresponding empirical distribution, i.e., a discrete probability distribution with support $ \mathcal{Y}_i = \{ \mathbf{y}_{i1}, \ldots, \mathbf{y}_{in_i} \}$, where we ignore any duplicate values in the definition of $\mathcal{Y}_i$, and $\delta_{\mathbf{y}_{il}}$ is the Dirac delta function at $\mathbf{y}_{il}$. 

We are interested in clustering the empirical distributions $\mu_1,\mu_2,\ldots,\mu_N$. Since there is no specific ordering among the distributions, we henceforth drop the subscript $i$ and denote the probability mass vector, support, and cardinality of the support of a generic empirical distribution $\mu$ as $\mathbf{p}_\mu$, $\mathcal{Y}_\mu$, and $M_\mu = |\mathcal{Y}_\mu|$, respectively. 

\subsection{Wasserstein Distance}
Let  $\Delta_{M} := \left\{\mathbf{p}\in \mathbb{R}_{+}^{{M}} \colon \sum_{l=1}^{{M} } p_l=1\right\}$ denote the set of all possible probability mass vectors on a support of size $M$, where $\mathbb{R}_{+}^{{M}}$ denotes the set of all real-valued length-$M$ vectors having only non-negative components.
For two empirical distributions $\mu$ and $\mu'$ having probability mass vectors $\mathbf{p}_\mu  \in \Delta_{M_\mu }$ and $\mathbf{p}_{\mu'} \in \Delta_{M_{\mu'}}$, respectively,  the polytope of couplings is defined as
$$
\Pi(\mathbf{p}_{\mu}, \mathbf{p}_{\mu'}) :=\left\{\boldsymbol{\gamma} \in \mathbb{R}_{+}^{M_{\mu} \times M_{\mu'}} \colon \boldsymbol{\gamma} \mathbf{1}_{M_{\mu'}}=\mathbf{p}_{\mu}, \boldsymbol{\gamma}^T \mathbf{1}_{M_{\mu}}=\mathbf{p}_{\mu'}\right\},
$$ where $\boldsymbol{\gamma}^T$ indicates the transpose of $\boldsymbol{\gamma}$, and $ \mathbf{1}_{M}$ indicates a length-$M$ vector of all $1$s. The polytope $\Pi(\mathbf{p}_{\mu}, \mathbf{p}_{\mu'})$ represents the set of all possible matrices $\boldsymbol{\gamma}$ that redistribute the probability mass from $\mathbf{p}_{\mu}$ to $\mathbf{p}_{\mu'}$, where each matrix entry $\boldsymbol{\gamma}_{ll'}$ represents the amount of mass transported from the $l$th element in $\mathcal{Y}_{\mu}$ to the $l'$th element in $\mathcal{Y}_{\mu'}$ for $l=1,2,\ldots,M_\mu$ and $l' = 1,2,\ldots, M_{\mu'}$, where the indexing of the supports is arbitrary. 
The Wasserstein distance between $\mu$ and $\mu'$, denoted by $W ({\mu},{\mu'})$, is defined as the optimal value of the following optimization problem:
\begin{equation}
\label{equ: wasserstein distance}
W ({\mu},{\mu'}) := \underset{\boldsymbol{\gamma} \in \Pi( \mathbf{p}_{\mu},\, \mathbf{p}_{\mu'} )}
{\operatorname{min}} \langle \mathbf{D},\boldsymbol{\gamma}\rangle,
\end{equation}
where $\mathbf{D} \in \mathbb{R}^{M_{\mu} \times M_{\mu'}}$ is a cost matrix consisting of the pairwise distances between points in $\mathcal{Y}_{\mu}$ and $\mathcal{Y}_{\mu'}$, and $\langle \cdot \, , \cdot \rangle$ denotes the summation of the element-wise product of two matrices. The optimal solution to the linear program posed in (\ref{equ: wasserstein distance}), denoted by $\boldsymbol{\gamma}^*$, is often referred to as the transportation plan matrix and represents the optimal allocation of probability mass from the source distribution $\mu$ to the target distribution $\mu'$. The time complexity of algorithms for computing $\boldsymbol{\gamma}^*$ is proportional to the cube of the support size, assuming that both distributions have the same support size \citep{altschuler2017near, cuturi2013sinkhorn}.

The regularized Wasserstein distance \citep{genevay2018learning}, on the other hand, can be computed in near-linear time \citep{cuturi2013sinkhorn} and is defined as 
\begin{equation}
\label{equ: regularized wasserstein distance}
 W_\lambda(\mu, \mu') :=  \langle \mathbf{D},  \boldsymbol{\boldsymbol{\gamma}_\lambda^*} \rangle ,\quad \text{where} \quad \boldsymbol{\gamma}_\lambda^* := \underset{\boldsymbol{\gamma}_\lambda \in \Pi(p_{\mu}, p_{\mu'})}{\operatorname{argmin}} \langle \mathbf{D}, \boldsymbol{\gamma}_\lambda \rangle - \lambda E(\boldsymbol{\gamma}_\lambda)
\end{equation}
where $\lambda > 0$ is a regularization parameter, and $E(\boldsymbol{\gamma}_\lambda)$ is the entropy of the transportation plan matrix $\boldsymbol{\gamma}_\lambda$, defined as $E(\boldsymbol{\gamma}) :=-\sum_{l=1}^{M_{\mu}} \sum_{l'=1}^{M_{\mu'}} \boldsymbol{\gamma}_{ll'} \log \boldsymbol{\gamma}_{ll'}$, where we set $\boldsymbol{\gamma}_{ll'} \log \boldsymbol{\gamma}_{ll'} = 0 $ if $\boldsymbol{\gamma}_{ll'} = 0$. 
As $\lambda$ approaches $0$, the optimal transportation plan matrix for (\ref{equ: regularized wasserstein distance}), $\boldsymbol{\gamma}_\lambda^*$, becomes more sparse and approaches $\boldsymbol{\gamma}^*$ \citep{benamou2015iterative}. The entropic regularization term incentivizes $\boldsymbol{\gamma}_\lambda^*$ to be more diffuse than $\boldsymbol{\gamma}^*$, and this induced non-sparsity helps to stabilize the computation of $\boldsymbol{\gamma}_\lambda^*$ because (\ref{equ: regularized wasserstein distance}) is a strongly convex program with a unique solution. An advantage of the regularized Wasserstein distance is that $\boldsymbol{\gamma}^*_\lambda$ can be calculated through an efficient iterative procedure involving matrix multiplications, as described in Algorithm~\ref{alg1:regularized Wasserstein Distance}.

\begin{algorithm}[htb]
\caption{Regularized Wasserstein distance \citep{benamou2015iterative}}
\label{alg1:regularized Wasserstein Distance}
\begin{algorithmic}[1]
\Statex \textbf{Input:} $\mu, \mu', \lambda,$ $\mathbf{D}$
\State Construct a matrix $\mathbf{Q} \in \mathbb{R}^{M_\mu \times M_{\mu'}}$ having entries $Q_{ll'} = e^{-\frac{\mathbf{D}_{ll'}}{\lambda}}$ for $l=1,2,\ldots,M_\mu$ and $l'=1,2,\ldots,M_{\mu'}$.
\State Initialize $\mathbf{v}^{(0)} = \mathbf{1}_{M_{\mu'}}$ and $m=0$.
\While{stopping criteria not met}

\State Set $\mathbf{u}^{(m)} = \frac{\mathbf{p}_\mu}{\mathbf{Q} \mathbf{v}^{(m)}}$, $\mathbf{v}^{(m+1)} = \frac{\mathbf{p}_{\mu'}}{\mathbf{Q}^T \mathbf{u}^{(m)}}$, and $\boldsymbol{\gamma}_\lambda^{(m)} = \operatorname{diag}(\mathbf{u}^{(m)}) \mathbf{Q} \operatorname{diag}(\mathbf{v}^{(m)}).$
\State $m \gets m + 1$.
\EndWhile
\State \textbf{return} $W_\lambda(\mu, \mu') = \langle \mathbf{D}, \boldsymbol{\gamma}_\lambda \rangle$.
\end{algorithmic}
\end{algorithm}

\begin{figure}[!htbp]
    \centering
    \includegraphics[width=0.5\textwidth]{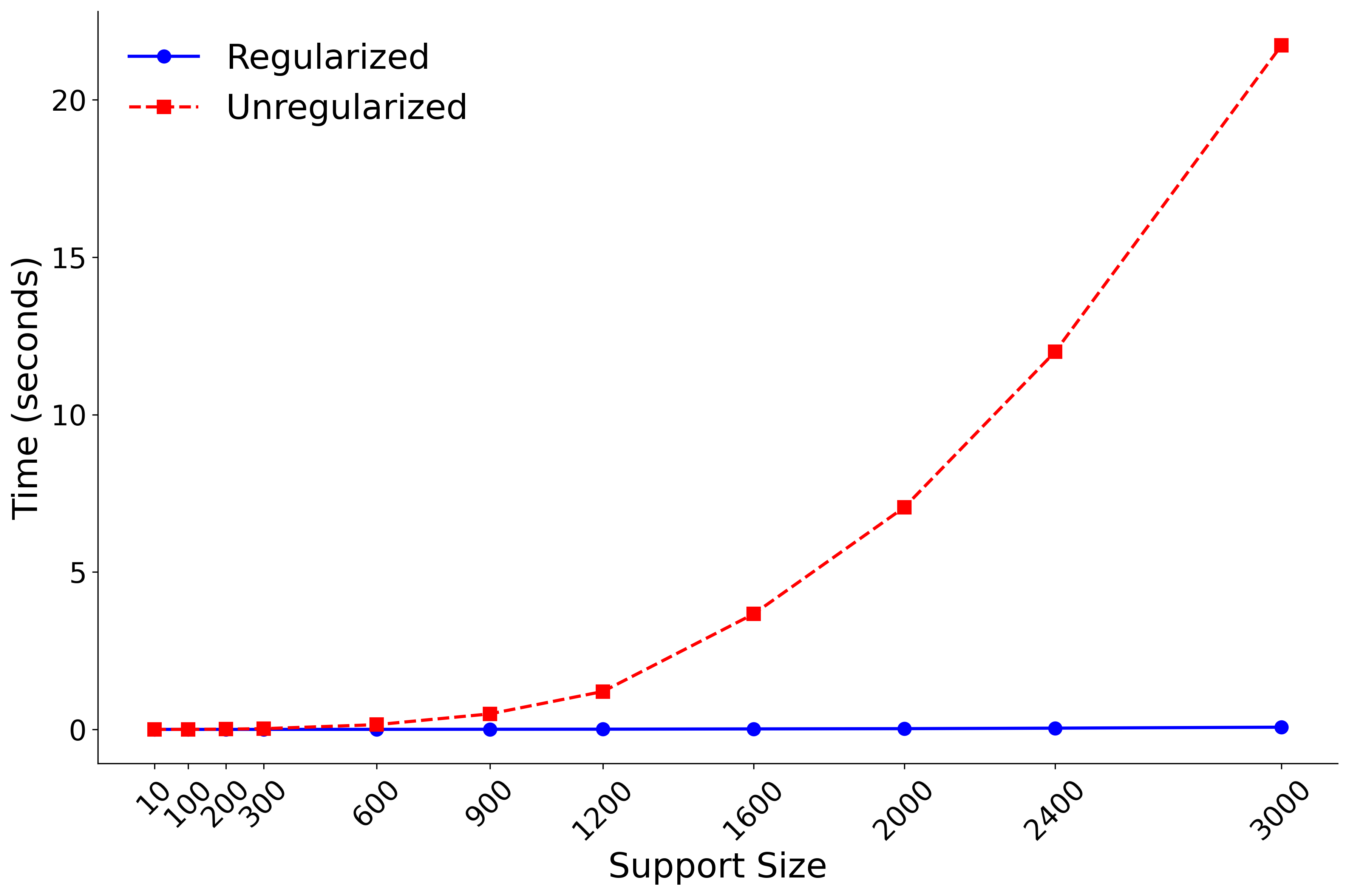} 

    \caption{Computational cost of calculating the Wasserstein distance and  regularized Wasserstein distance between two samples generated with the same support size, varying between 1 and 3000, with calculations repeated 10 times for each support size. The variability was negligible compared to the difference in computational costs between the two distance measures.}
    \label{fig: reg vs unreg}
\end{figure}

The stopping criteria in Algorithm \ref{alg1:regularized Wasserstein Distance} helps to control the computational cost and could involve setting a maximum number of iterations or stopping when the percentage change in the regularized Wasserstein distance is less than some threshold. The regularized Wasserstein distance plays a central role in our proposed clustering algorithm.
Figure \ref{fig: reg vs unreg} demonstrates that the time required to calculate the Wasserstein distance increases exponentially in the support size, while the cost of computing the regularized Wasserstein distance increases nearly linearly.

\subsection{An Agglomerative Clustering Algorithm}
Agglomerative clustering is a hierarchical clustering method that begins by treating each instance as an individual cluster and successively merges the closest pairs based on a specified distance metric, allowing clusters to form organically from the data.
We choose to employ agglomerative clustering for several reasons. Firstly, unlike $k$-means clustering, in which the number of clusters is predefined, agglomerative clustering is a better choice in situations where the optimal number of clusters is unknown. 
Secondly, centroid-based methods, such as $k$-means, are sensitive to outliers due to their reliance on a single central point to represent each cluster. Outliers can significantly skew the centroid's location and distort the clustering process. In contrast, the complete-linkage approach commonly used in agglomerative clustering considers the maximum distance between any two points in distinct clusters, making the clustering more robust to outliers. Complete-linkage clustering also considers the farthest points within the merged clusters, resulting in tighter and more spherical cluster formations compared to single-linkage clustering, which can generate elongated clusters due to chaining effects. 
Thirdly, agglomerative clustering, particularly when employing complete linkage, offers a valuable output in the form of a dendrogram, which depicts the merging process and the distances between clusters at each stage of the algorithm and can aid in comprehending the relationships between instances and in determining an appropriate number of clusters.
Fourthly, unlike the $k$-means algorithm in which one must repeatedly recalculate the centroid of each cluster, agglomerative clustering does not entail calculating centroids. 

To determine the optimal number of clusters, we use the silhouette index proposed by \cite{shahapure2020cluster}. The silhouette index for a given clustering $\mathcal{C}$ is defined as
$$
S_{\mathcal{C}}=\frac{1}{|\mathcal{C}|} \sum_{C \in \mathcal{C}} \frac{1}{|C|} \sum_{\mu \in C} S_{\mu} 
,$$ where $S_{\mu} =( b_{\mu} - a_{\mu} )/ \max\{b_{\mu}, a_{\mu}\}$,
$$a_{\mu} = (|C_{\mu}|- 1)^{-1}\sum_{\mu' \in C_{\mu},\, \mu' \neq \mu} W_\lambda(\mu, \mu')
$$
is the average regularized Wasserstein distance between $\mu$ and every other distribution in the same cluster, and
$$
b_{\mu} = \min_{C' \in \mathcal{C},\, C' \neq C_{\mu}}\left\{(|C'|-1)^{-1} \sum_{\mu' \in C'} W_\lambda(\mu, \mu') \right\}
$$
is the minimum average distance between $\mu$ and distributions in other clusters. The silhouette index considers both intra-cluster (as in $a_\mu$) and inter-cluster (as in $b_\mu$) distances, and its values fall within the range of $-1$ to $1$, with a higher silhouette index indicating a more favorable clustering. For an individual distribution $\mu$, a silhouette index $S_\mu$ close to 1 signifies that $\mu$ is well positioned within its assigned cluster.

We now present Algorithm~\ref{alg4: Hierarchical Clustering}, an agglomerative algorithm for clustering the multivariate empirical distributions of simulation outputs.
In Algorithm~\ref{alg4: Hierarchical Clustering},  $\mathbf{D}$ denotes the cost matrix between points in the supports. Before applying Algorithm \ref{alg4: Hierarchical Clustering}, the output data is normalized within each dimension to ensure that no one KPI skews the clustering results.

\begin{algorithm}[htb]
\caption{Hierarchical clustering with regularized Wasserstein distance}
\label{alg4: Hierarchical Clustering}
\begin{algorithmic}[1]
\Statex \textbf{Input:} $\{\mu_1,\ldots,\mu_N\}$, $\lambda$,  $\mathbf{D}$
\State Compute the regularized Wasserstein distance between all pairs of empirical distributions.
\State Initialize $\mathcal{C}$ as the clustering where each distribution is in its own cluster.
\While{$|\mathcal{C}| > 1$}
    \State Identify the two clusters $C^*$ and $C'^*$ that are closest to each other based on the complete-linkage distance calculation rule, i.e., $\underset{C,C' \in \mathcal{C}}{\operatorname{argmin}} \underset{\mu \in C , \mu' \in C'}{\operatorname{max}} W_\lambda(\mu, \mu').$
    \State Merge clusters $C^*$ and $C'^*$ and update $\mathcal{C}$. 
    \State Compute the silhouette index for $\mathcal{C}$. 
\EndWhile
\State Choose the clustering with the largest silhouette index.
\end{algorithmic}
\end{algorithm}

To further assess the practicality of Algorithm \ref{alg4: Hierarchical Clustering}, it is essential to consider its computational cost. Step 1 of Algorithm \ref{alg4: Hierarchical Clustering} calculates the pairwise distances between all distributions, the cost of which scales quadratically with the number of distributions, $N$. Calculating the regularized Wasserstein distance between a pair of distributions with the same support size exhibits a quadratic dependence on the size of the support \citep{altschuler2017near}. After obtaining the pairwise distances, the rest of Algorithm~\ref{alg4: Hierarchical Clustering} has a cubic dependence on $N$ \citep{karthikeyan2020comparative}. Additionally, calculating the silhouette index for a given clustering scales quadratically with $N$ \citep{Mur2016}.

\textbf{Remark} In experiments with stochastic simulation models, the user has complete control over how pseudo-random numbers are used to drive the simulation model.
One such scheme is \textit{common random numbers} (CRN), which refers to the technique of using the same streams of pseudo-random numbers when simulating different scenarios in an attempt to induce positive correlation in the outputs on any given replication \citep{KELTON2006181}.
When successful, CRN enables more accurate comparisons between scenarios and is a powerful tool for simulation optimization.
We believe that CRN can have benefits for clustering as well. In particular, we conjecture that CRN will tend to reduce the variance in the Wasserstein distances between output distributions and consequently reduce the variance of the clustering outcomes. 
Furthermore, we anticipate that CRN should yield clusterings that more closely resemble the true underlying clustering structure, i.e., how the scenarios' true outputs distributions would be clustered were they known. 
As we do not believe these benefits of CRN are universal, we eschew attempting to prove theoretical statements about them and instead provide some empirical support for these assertions in Section~\ref{sec:experiments}.

\subsection{Wasserstein Barycenter}
After clustering the distributions, we turn to the regularized Wasserstein barycenter to summarize the information in each cluster. For a given cluster, the regularized Wasserstein barycenter minimizes the average regularized Wasserstein distance between itself and each of the distributions within the cluster, effectively acting as an ``average'' of the distributions. There are several approaches to compute the barycenter for a cluster $C$, denoted generically by $\bar{\mu}$. One approach is to fix the support and then optimize over the probabilities assigned to each support point. Alternatively, we could fix the number of support points and their corresponding probabilities, as discussed in \cite{altschuler2017near}, and then optimize over the locations of the support points. We choose to  employ an iterative algorithm introduced in \cite{cuturi2014fast} to simultaneously optimize both the support points and their corresponding probabilities, with the only fixed parameter being the support size. This procedure continues until the stopping criterion, defined similarly to Algorithm \ref{alg1:regularized Wasserstein Distance}, is met. 
Specifically we let the support of the barycenter be $\mathcal{Y}_{\bar{\mu}}$. The regularized Wasserstein barycenter of a cluster $C$ is a discrete distribution on $\mathcal{Y}_{\bar{\mu}}$ having a probability mass vector
$$
\mathbf{p}_{\bar{\mu}} := \underset{\mathbf{p} \in \Delta_{M_{\bar{\mu}}}}{\operatorname{argmin}} \frac{1}{|C|} \sum_{\mu \in C} W_{\lambda}(\mu, \bar{\mu}).
$$
In this paper, we assume that the scenarios are equally important, and hence equally weighted, but the Wasserstein barycenter can be derived by minimizing a weighted sum of regularized Wasserstein distances
The optimal probability mass vector $\mathbf{p}_{\bar{\mu}}$ for a fixed support can be computed using another iterative procedure, given in Algorithm \ref{alg: opt prob}. Then by incorporating Algorithm \ref{alg: opt prob} into Algorithm \ref{alg: reg barycenter}, we can allow the support to be free. Although we have regarded the supports as sets, in Algorithm~{alg: reg barycenter} we will denote the supports $\mathcal{Y}_{\bar{\mu}}$ and $\mathcal{Y}_{\mu}$ as matrices $\boldsymbol{Y_{\bar{\mu}}}$ and $\boldsymbol{Y_\mu}$, with the understanding that the ordering of elements within these matrices is arbitrary. The parameters $\beta$ and $\theta$ appearing in Algorithms~\ref{alg: opt prob} and~\ref{alg: reg barycenter} vary between 0 and 1 and control how the step sizes are updated.

\begin{algorithm}
\caption{Finding the optimal $p_{\bar{\mu}}$ for a fixed support \citep{cuturi2014fast}}
\label{alg: opt prob}
\begin{algorithmic}[1]
\Statex \textbf{Input:} C, $\lambda$, $\mathbf{D}$, $\beta$
\State Set $\hat{p}_{\bar{\mu}} = \tilde{p}_{\bar{\mu}} = \mathbf{1}/{|\mathcal{Y}_{\bar{\mu}}|}$.
\While{not converged}
    \State $\beta = (t + 1)/2$, $p_{\bar{\mu}} = (1 - \beta^{-1}) \hat{p}_{\bar{\mu}} + \beta^{-1} \tilde{p}_{\bar{\mu}}$.
    \State Calculate $\alpha_{\bar{\mu}} = \frac{1}{|C|} \sum_{\mu \in C} \alpha_\mu$ where $\alpha_\mu = -\lambda\log(u_\mu)$ is derived when finding the regularized Wasserstein distance between $\mu$ and $\bar{\mu}$ using Algorithm \ref{alg1:regularized Wasserstein Distance}.
    \State $\tilde{p}_{\bar{\mu}} \leftarrow \tilde{p}_{\bar{\mu}} \odot e^{-t_0\beta \alpha_{\bar{\mu}}}$, $\tilde{p}_{\bar{\mu}} \leftarrow \frac{\tilde{p}_{\bar{\mu}}}{\tilde{p}_{\bar{\mu}}^T \mathbf{1}_{|C|}}$.
    \State $\hat{p}_{\bar{\mu}} \leftarrow (1-\beta^{-1}) \hat{p}_{\bar{\mu}} + \beta^{-1} \tilde{p}_{\bar{\mu}}$, $t \leftarrow t+1$.
\EndWhile
\State \textbf{return} $p_{\bar{\mu}}$.
\end{algorithmic}
\end{algorithm}

\begin{algorithm}
\caption{Regularized Wasserstein barycenter \citep{cuturi2014fast}}
\label{alg: reg barycenter}
\begin{algorithmic}[1]
\Statex \textbf{Input:} C, $\lambda$, $\mathbf{D}$, $\beta$, $\theta$.
\State Initialize $\boldsymbol{Y}_{\bar{\mu}}$ and  $p_{\bar{\mu}}$.
\While{$\mathcal{Y}_{\bar{\mu}}$ and $p_{\bar{\mu}}$ have not converged}
    \State Derive $p_{\bar{\mu}}$ using Algorithm \ref{alg: opt prob}.
    \For{$\mu \in C$}
        \State Obtain $\gamma_{\lambda,\mu} $ using $W_\lambda(\bar{\mu}, \mu)$.
    \EndFor
    \State $\boldsymbol{Y}_{\bar{\mu}} = (1 - \theta)\boldsymbol{Y}_{\bar{\mu}} + \theta \left( \frac{1}{|C|} \sum_{\mu \in C} \boldsymbol{Y}_{\mu}  \gamma_{\lambda,\mu }\right) \text{diag}(p_{\bar{\mu}})^{-1}$.
\EndWhile
\State \textbf{return} $\boldsymbol{Y}_{\bar{\mu}}$, $p_{\bar{\mu}}$.
\end{algorithmic}
\end{algorithm}

\begin{figure}[!htbp]
    \begin{subfigure}[b]{0.49\textwidth}
        \centering
        \includegraphics[width=\textwidth]{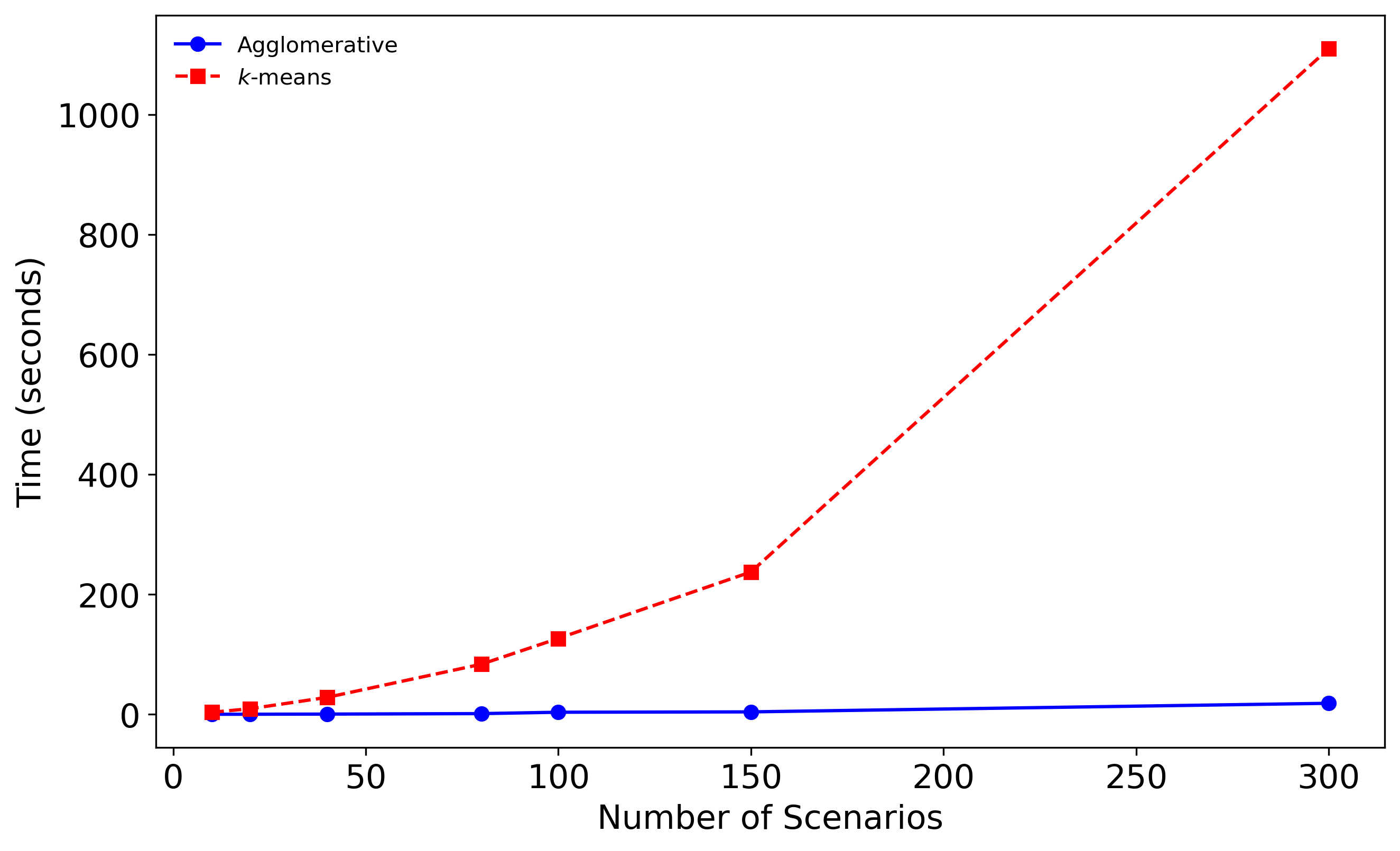}
        \label{fig: comparison_num_of_systems}
    \end{subfigure}
    \hfill 
    \begin{subfigure}[b]{0.49\textwidth}
        \centering
        \includegraphics[width=\textwidth]{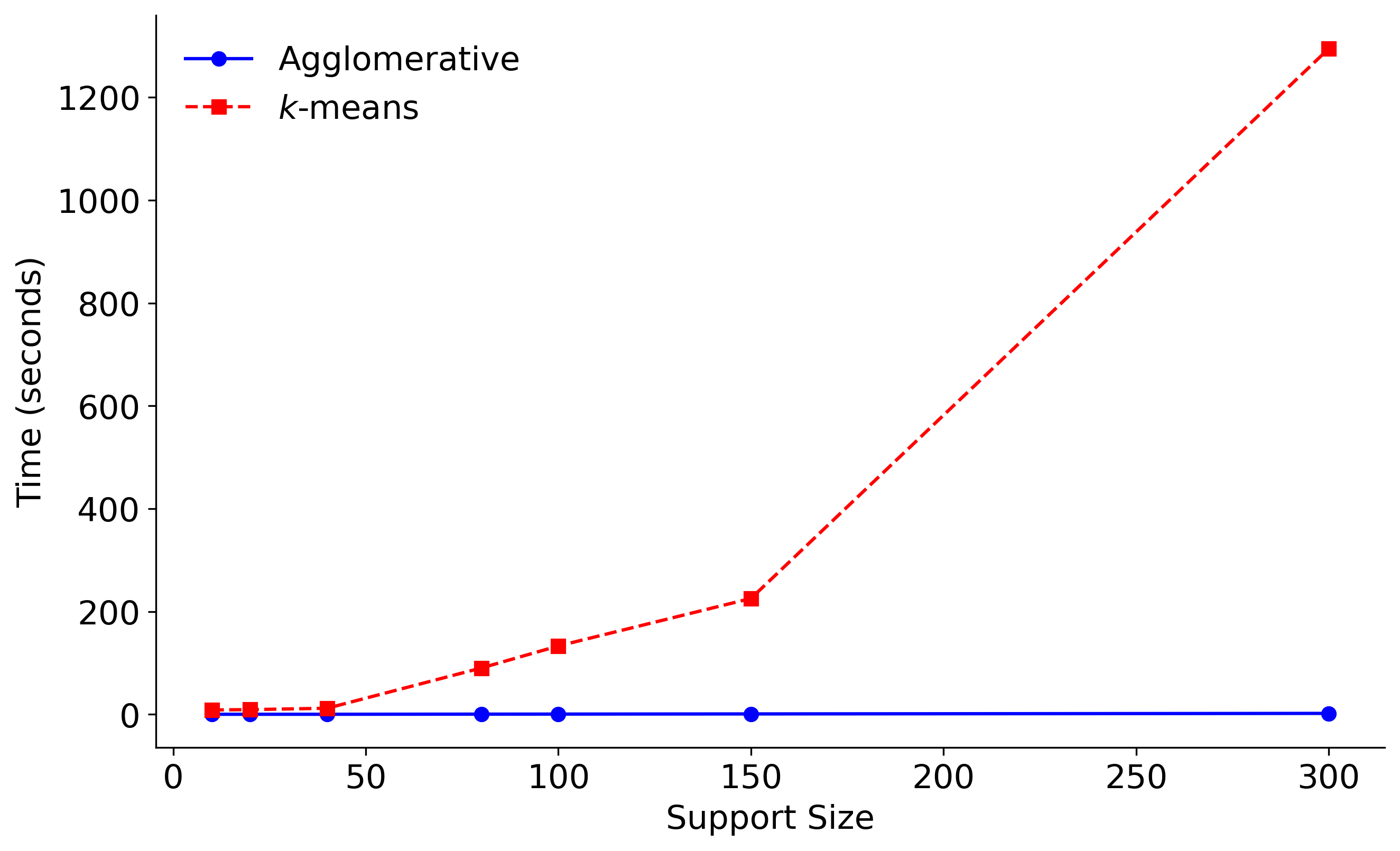}
        \label{fig: comparison_size_of_support}
    \end{subfigure}
    \caption{ Comparison of computational times for $k$-means and agglomerative clustering for varying numbers of distributions with a fixed support size of 20 (Left) and for varying support sizes for 20 distributions (Right). The variability was negligible compared to the difference in computational costs between the two algorithms. }
    \label{fig:combined_correlation}
\end{figure}

The Wasserstein barycenter lacks robustness \citep{santambrogio2016convexity}, and this idiosyncratic behavior makes centroid-based clustering approaches like \( k \)-means inadequate for representing inter-cluster variability. Additionally, agglomerative clustering is more efficient in terms of computational cost. After introducing the barycenter concept, we compare the efficiency of our hierarchical algorithm with that of a \( k \)-means algorithm using barycenters defined in Algorithm~\ref{alg: reg barycenter} for the cluster centroids. In Figure~\ref{fig:combined_correlation}, we fix the number of scenarios under consideration and vary the number of replications from 10 to 300, common across scenarios. In a second experiment, we fix the support size---the number of replications---and vary the number of scenarios from 10 to 300. For the $k$-means algorithm, we do not know the optimal number of clusters and therefore we choose to calculate the clusterings for 2 to 10 clusters. In Figure~\ref{fig:combined_correlation}, we observe that the computational cost increases with the number of scenarios for both $k$-means and agglomerative clustering, however, the rate of increase is significantly smaller for agglomerative clustering. The relatively high computational cost of the $k$-means algorithm is due to the repeated calculation of barycenters. The number of such calculations depends on the number of systems and clusters, whereas the time for each calculation depends on the support size of the distributions and the barycenter. In this experiment, the support size of the barycenter is taken to be the same as that of the input distributions.

\section{Experiments}
\label{sec:experiments}

We demonstrate several use cases of the proposed clustering algorithm and several ways to plot the results through experiments involving a discrete-event simulation model of a call center. The call center operates from 8am to 4pm and during this time customers call in according to a stationary Poisson process with a rate of 400 customers per hour. This call center serves two classes of customers---regular and premium---with regular customers comprising 60$\%$ of incoming calls. Two sets of operators---basic service and premium service---provide initial service to regular and premium customers, respectively. If there are no premium customers in the queue, premium service operators can serve regular customers; however, basic service operators cannot serve premium customers. Additionally, 15$\%$ of arriving customers, irrespective of their class, abandon if their initial service does not start within a customer-specific amount of time following a uniform distribution between 0.5 and 3 minutes.
After their initial service is completed, 15$\%$ of customers, irrespective of their class, require additional service that is provided by a third type of operator: technical. Regular and premium customers are served by the same team of technical operators.  Service times for basic service,  premium service, and technical operators follow exponential distributions with means of 7, 3, and 10 minutes, respectively. Operator-dependent service rates such as these may arise because premium service operators have more resources, full system access, and extensive experience, and therefore can resolve issues more quickly.
When queueing for technical support, premium customers are given priority over regular customers, and customers do not abandon. The call center stops receiving new calls at the end of the workday but continues operating until all customers have been served; this policy imposes overwork on the operators.

\subsection{Staffing a Fixed Number of Operators}

\begin{figure}[htb]
\captionsetup{skip=-10pt}
    \centering
    \begin{subfigure}[b]{0.49\textwidth}
        \includegraphics[width=\textwidth]{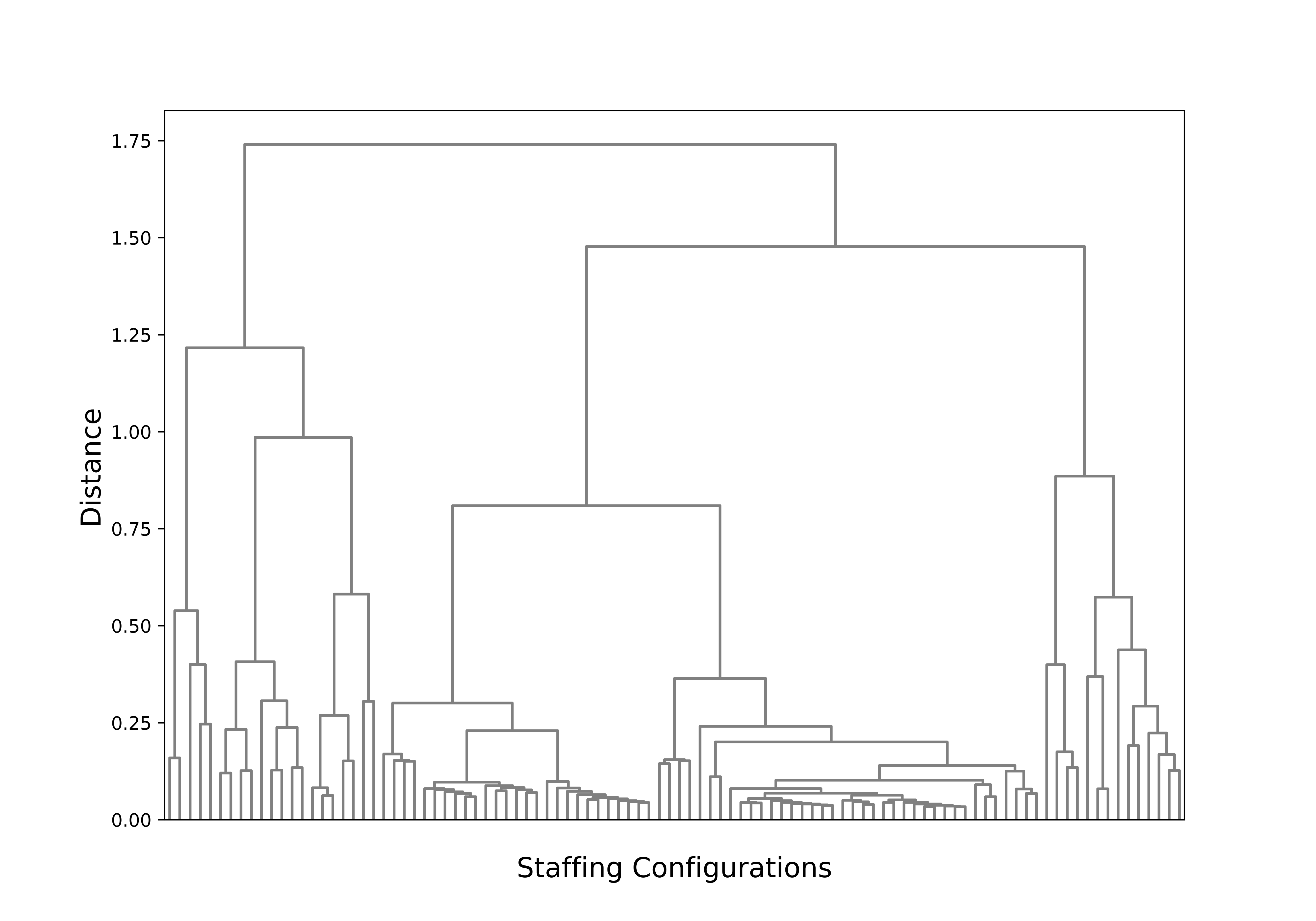}
        \label{fig: plot_dendrogram}
    \end{subfigure}
    \hfill 
    \begin{subfigure}[b]{0.49\textwidth}
        \includegraphics[width=\textwidth]{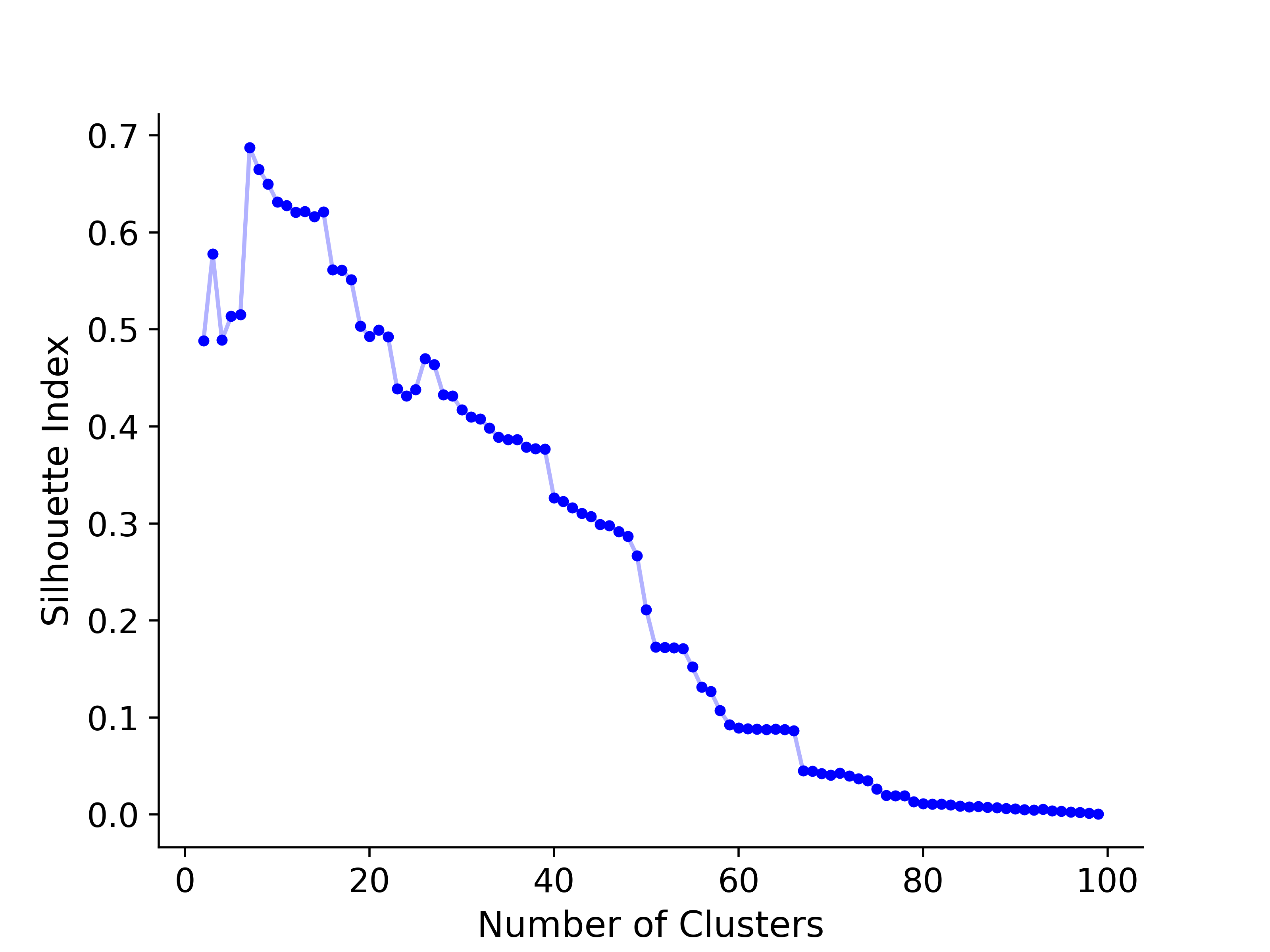}
        \label{fig: plot_silhouette_optimization}
    \end{subfigure}
    \caption{(Left) Dendrogram from clustering the output distributions of 100 staffing configurations. (Right) Silhouette index for different clusterings produced by Algorithm \ref{alg4: Hierarchical Clustering}.}
    \label{fig:combined}
\end{figure}

Suppose the call-center manager needs to train 49 operators for some combination of basic service, premium service, and technical roles and is interested in five KPIs: the mean time in the system for regular (${Y_1}$) and premium customers (${Y_2}$), and the mean overwork time for basic service (${Y_3}$), premium service ($Y_4$), and technical operators (${Y_5}$). Assuming that there must be at least one operator of each type, there are 1128 possible staffing configurations (scenarios). We show that even when simulating a fraction of these configurations, clustering can provide valuable insights about the system's behavior. 
We choose 100 configurations that uniformly cover the space of all configurations and simulate 40 days (replications) under each configuration. We then apply Algorithm \ref{alg4: Hierarchical Clustering} to cluster the obtained empirical distributions for the five KPIs. The dendrogram in Figure~\ref{fig:combined} shows the hierarchical clustering of the simulated configurations. Based on the silhouette index plot shown in Figure~\ref{fig:combined}, having 7 clusters is a good choice, though having 8 or 9 clusters would also be satisfactory.

\begin{figure}[!htbp]
    \centering
    \includegraphics[width=0.90\textwidth]{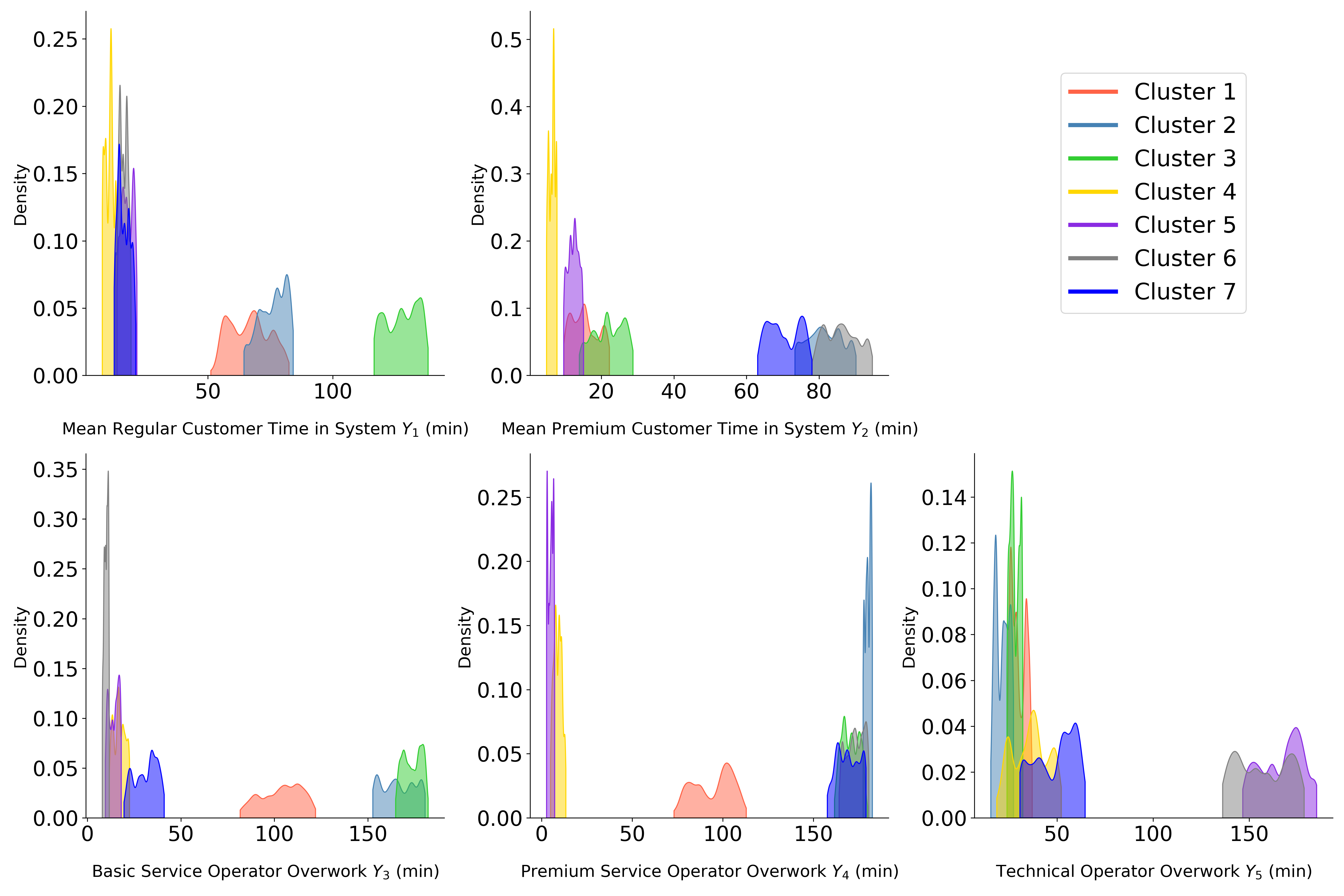} 
    \caption{Marginal pdfs for each KPI for each barycenter, for a fixed number of operators.}
    \label{fig: plot_ecdf}
\end{figure}

To more deeply understand the distributions within each cluster, we compute the barycenter of each cluster and plot the kernel density estimates of probability density functions (pdfs) for each of the five KPIs. In Figure \ref{fig: plot_ecdf}, we observe that no cluster consistently outperforms the others; however, Cluster 4 performs well across all five KPIs, whereas the other clusters perform poorly in at least one KPI.

\begin{figure}[!htb]
        \centering
        \includegraphics[scale=0.55]{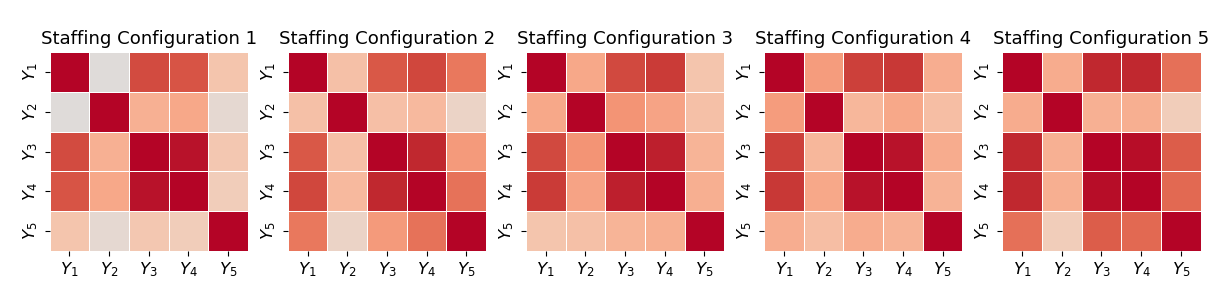}

        \caption{Correlation matrices for staffing configurations in Cluster 1. }
        
        \label{fig: correlation_matrix_cluster_1}

\end{figure}

Another cluster of interest is Cluster 1, as it performs well in two KPIs and moderately (relative to other clusters) in the remaining three KPIs. To understand this behavior better, we can examine the correlation matrices of the distributions in this cluster, depicted in Figure \ref{fig: correlation_matrix_cluster_1}. The matrices reveal strong positive correlations between the mean regular customer waiting time, $Y_1$, the overwork of basic operators, $Y_3$, and the overwork of premium operators, $Y_4$. This suggests that on days for which the regular customers' waiting time is above average, the workload increased for both types of operators, and on any given day observing one of these correlated metrics can potentially reveal the trend of the others. 

\begin{figure}[!htb]
        \centering
        \includegraphics[scale=0.6]{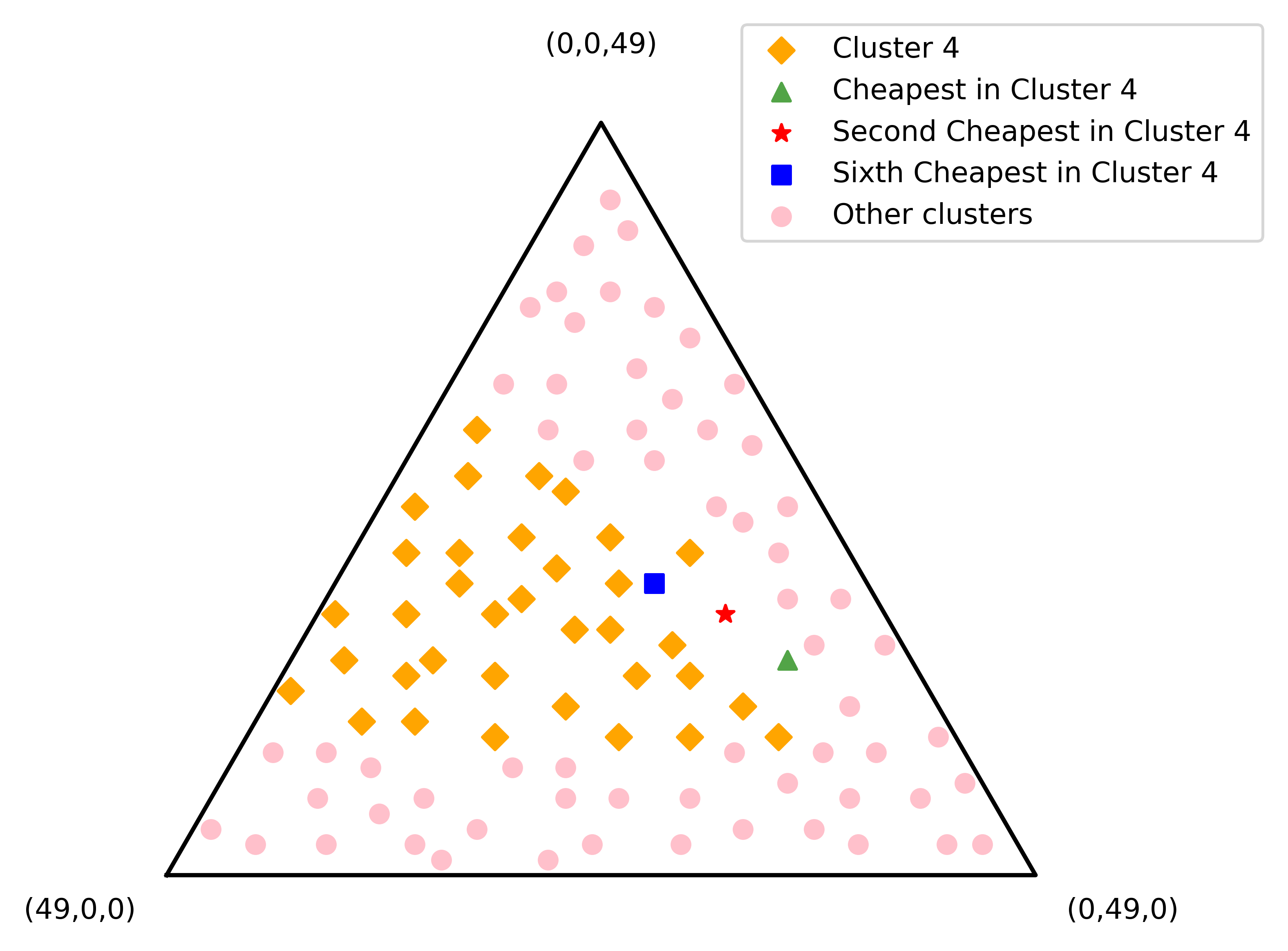}
  
        \caption{Simulated staffing configurations, represented as triplets of (\# of premium service operators, \# of basic service operators, \# of technical operators). Configurations in Cluster 4, including the cheapest (7, 28, 14), second cheapest (9, 23, 17), and sixth cheapest (12, 18, 19) configurations in Cluster 4, are highlighted.}
        
        \label{fig: plot_input_3d}

\end{figure}

\begin{figure}[!htb]
        \centering
        \includegraphics[width=\textwidth]{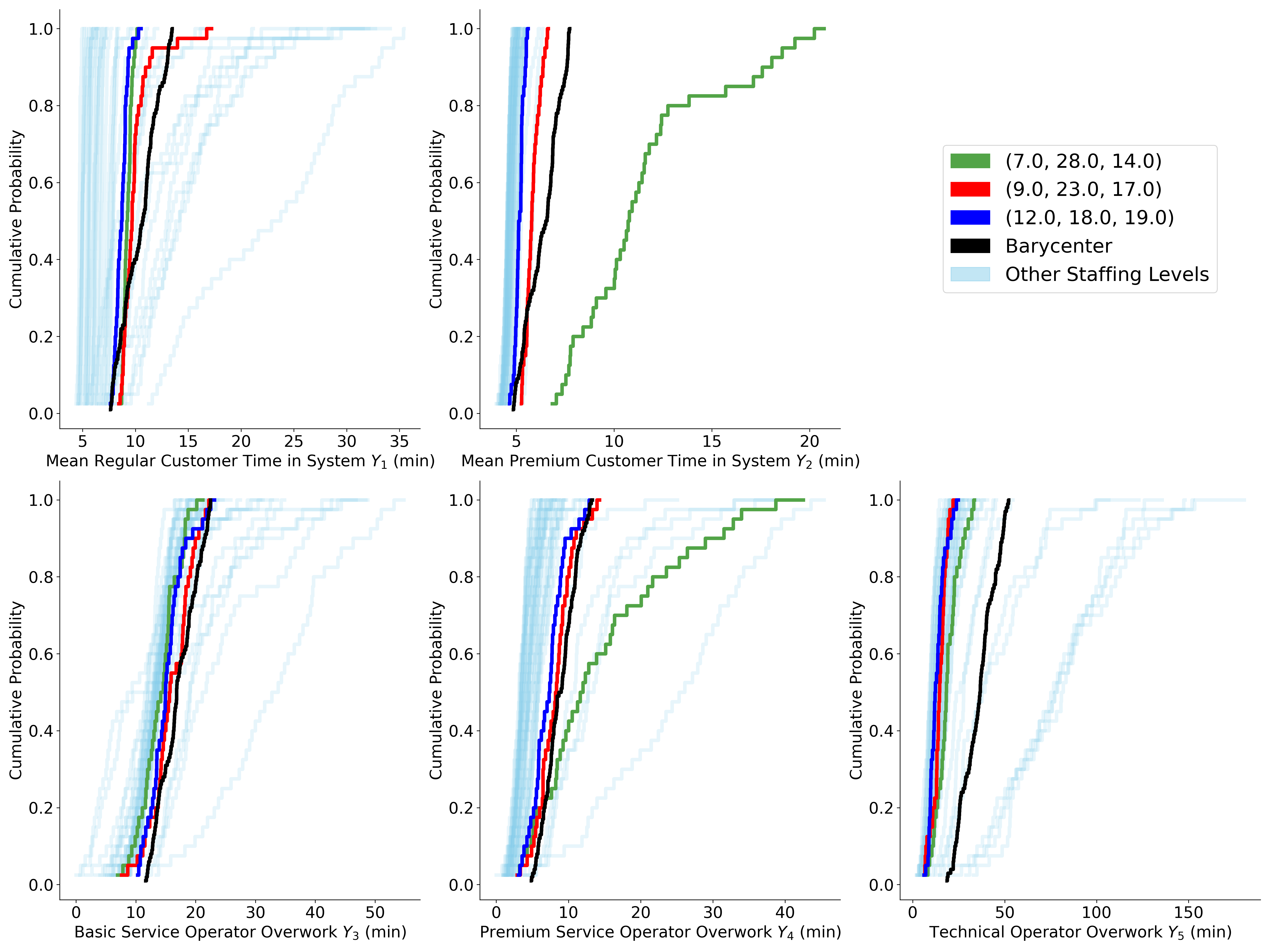}
        \caption{Comparison of marginal cdfs for different staffing configurations in Cluster 4.}
        
        \label{fig: best ones}

\end{figure}

Returning to our analysis of Cluster 4, it can be seen in Figure~\ref{fig: plot_input_3d} that the staffing configurations that comprise Cluster 4 are characterized by having a moderate number of premium and technical operators and a variable number of basic service operators ranging from 1 to 30. The decision maker might want to identify the staffing configuration in Cluster 4 that, say, minimizes the total staffing costs.
For instance, if the staffing costs for basic service, premium service, and technical operators were 4, 1, and 1, respectively, then the configuration with 7 specialist, 28 regular, and 14 technical operators, would be the cheapest. Rather than choose staffing configuration (7, 28, 14) on the basis of cost alone, we recommend further inspecting the individual distributions within Cluster 4,  as barycenters predominantly capture central tendencies and may conceal significant intra-cluster variability or outliers.
The marginal cdfs of the configurations in Cluster 4, including that of its barycenter, are illustrated in Figure~\ref{fig: best ones}.  Upon further evaluation, we observe that staffing configuration $(7, 28, 14)$ performs adequately across $Y_1$, $Y_4$, and $Y_5$, however, its performance in $Y_2$ and $Y_3$ is unsatisfactory.  If we were to sort the staffing configurations in Cluster 4 by their costs and continue down the list, we would find that the second cheapest configuration in Cluster 4, (9, 23, 17), demonstrates solid performance across all metrics. However, staffing configuration (9, 23, 17) slightly underperforms the barycenter with respect to $Y_1$. If we were to insist on identifying a staffing configuration that stochastically dominates the barycenter in every performance, which many not necessarily exist, we would find that the sixth cheapest configuration, (12, 18, 19), is the cheapest configuration that does so. 

\subsection{Staffing Subject to a Budget}

\begin{figure}[htb]
    \centering
    \includegraphics[width=\textwidth]{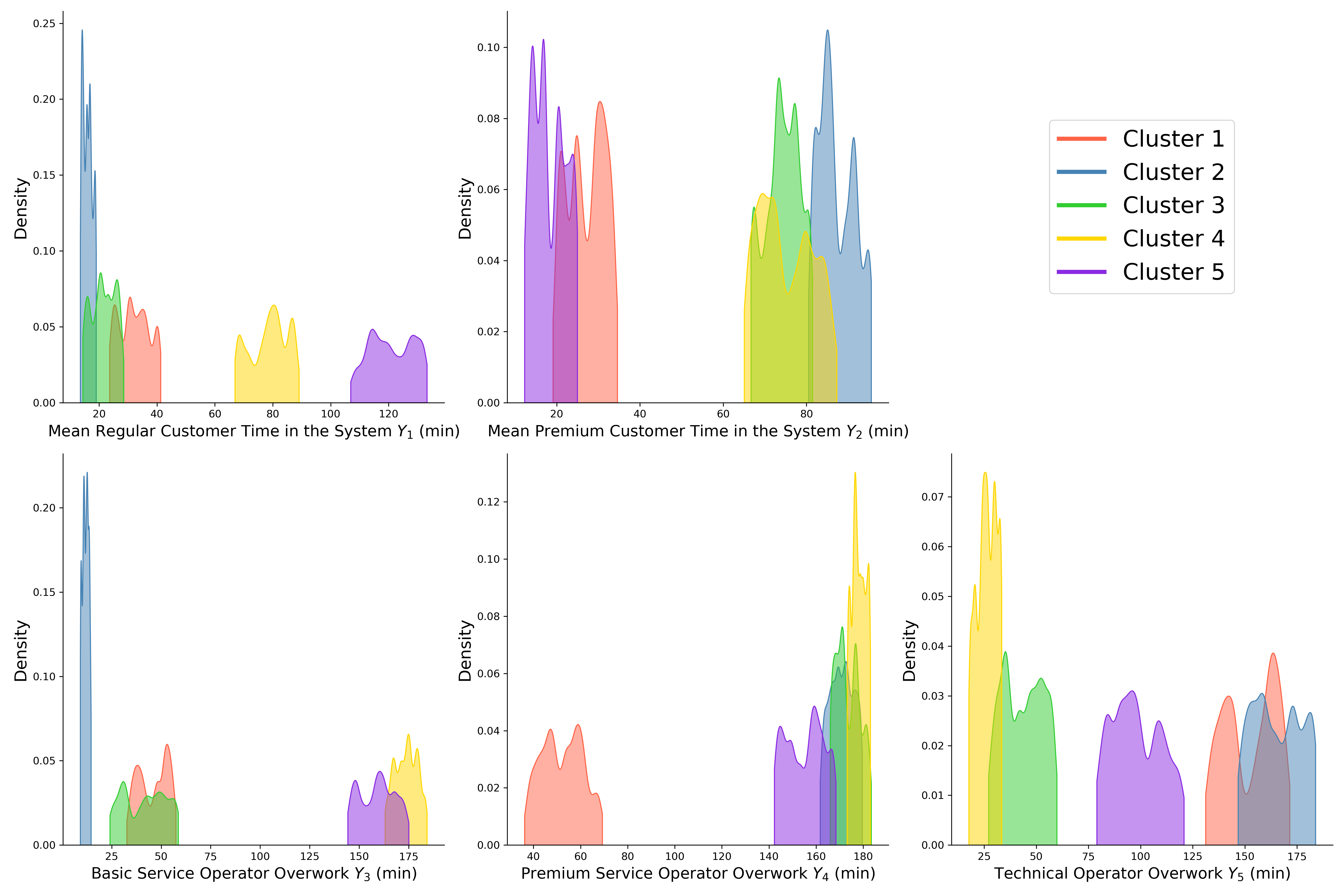} 
    \caption{Marginal pdfs for each KPI for each barycenter for staffing subject to a budget.}
    \label{fig: plot_ecdf_cost}
\end{figure}

Staffing-cost considerations could alternatively be incorporated into the design of the experiment. As a follow-up to the staffing-cost analysis above, suppose the decision maker's objective were to find a staffing configuration with a desirable output distribution among those having total costs between 50 and 55, assuming the same per-operator costs as before.
There are 2,143 feasible staffing configurations, and we again uniformly select 100 configurations, simulate each for 40 days, and apply Algorithm \ref{alg4: Hierarchical Clustering} on the results. The silhouette index recommends five clusters, and smoothed versions of the marginal pdfs of the corresponding barycenters are shown in Figure~\ref{fig: plot_ecdf_cost}.
Unlike in the previous experiment, no cluster dominates across all five KPIs: each cluster performs well in at least one KPI, but suffers in other aspects. The decision maker's priorities thus play a crucial role in balancing the tradeoffs across KPIs. For instance, if providing good customer service is more important than ensuring favorable conditions for operators, then Cluster 1 is preferable. Conversely, if keeping basic or technical operators' overwork low is a priority, then Cluster 3 might be preferred. Further investigation by the decision maker can identify the most cost-effective option or the configuration that outperforms other configurations in certain KPIs.

\subsection{Monitoring Queue Lengths}
\begin{figure}[htb]
    \centering
    \includegraphics[width=0.95\textwidth]{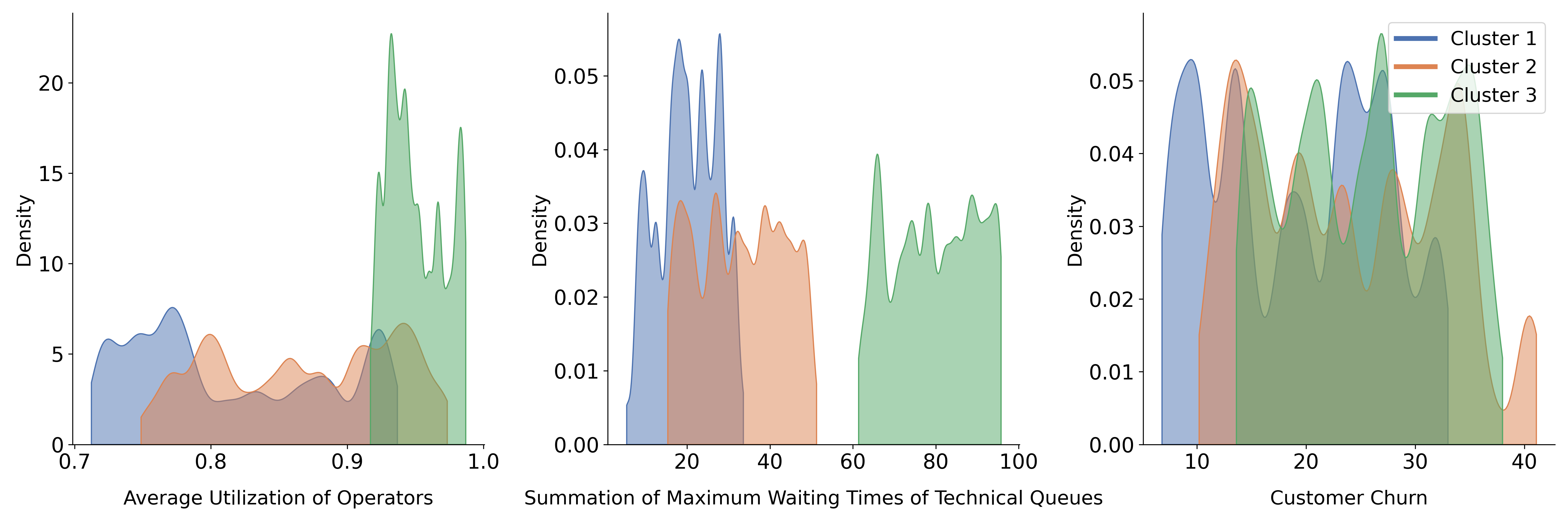} 
    \caption{Smoothed pdfs for each KPI for each barycenter for state online monitoring.}
    \label{fig: plot_ecdf_state_monitoring}
\end{figure}

In this experiment, we illustrate an approach that enables a decision maker to monitor the system in real time and use offline clustering to make staffing adjustments. In our setup, the system consists of 22 basic service, 9 premium service, and 8 technical operators, with other system specifications unchanged. The state of the system is represented by a 4-dimensional vector consisting of the queue lengths for regular and premium customers for initial service, and the queue lengths for regular and premium customers for technical service.
For a given state, we consider three KPIs: the mean utilization of all operators, the sum of the maximum waiting times of regular and premium customers in the technical queue, and the total number of customers who abandoned the queues (referred to as customer churn), all measured over a one-hour period when starting in that state. To construct a set of scenarios for our offline experiment, we first simulate the system for 5000 days, recording the states at the beginning of each hour along with the corresponding output vector after an hour of observation.
We restrict our attention to those states that were observed 10 or more times, of which there were 113. Algorithm \ref{alg4: Hierarchical Clustering} groups the output distributions into three clusters, the barycenters of which are depicted in Figure \ref{fig: plot_ecdf_state_monitoring}. Across all three performance metrics, Cluster 1 performs the best, Cluster 2 performs moderately well, and Cluster 3 performs the worst.

We could subsequently monitor and classify states visited during a new day by considering the state's two nearest neighbors, as measured in terms of the queue lengths. When the current state's two nearest neighbors belong to Cluster 1, we anticipate good performance in the next hour. Conversely, having both nearest neighbors in Cluster 2 suggests high customer churn and moderate operator utilization, with minimal impact on maximum waiting times. If both nearest neighbors belong to Cluster 3, poor performance across all metrics is expected in the next hour.
There will be cases where the nearest neighbors are of different kinds, i.e., we find ourselves in a transition state between clusters. In such cases, the decision maker should closely monitor trends and be prepared to take preventive actions, such as adjusting staffing in bottleneck areas or reallocating roles among cross-trained operators in a call-center context.

\begin{figure}[htb]
    \centering
    \includegraphics[width=\textwidth]{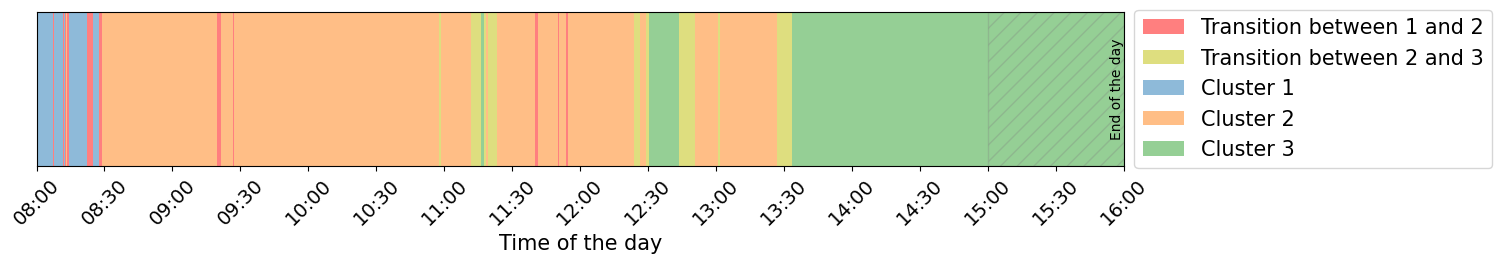}
    \caption{State monitoring of a new day with clustering and classification.}
    \label{fig: plot_state_monitoring}
\end{figure}

Figure \ref{fig: plot_state_monitoring} illustrates this monitoring approach over the course of one day. 
The system starts the day with empty queues (a good state), but before long it begins to oscillate between good and moderate states, before settling into a moderate state around 8:30 and remaining there until 11:15 with occasional transitions. Around 12:30, the system briefly shifts to a bad state, before returning to a moderate state until about 13:30, after which all states are bad. The dashed area represents the times when the call center will be closed within the next hour, but the state classifications during this period can still be useful.
The plot suggests several times where preventive action could be taken, e.g., around 11:20, when the system first enters an estimated bad state. A risk-averse decision maker might take preventive action at this time, but if they had waited to see if the situation persisted, they would have discovered that the system recovered on its own. 
Alternatively, the decision maker could intervene after observing bad states for some given duration. Each approach has its merits, catering to different risk tolerances and operational strategies.

\subsection{Common Random Numbers}
In our final experiment, we investigate the effects of employing CRN on the clustering process.
To implement CRN in the call center model, we modify the service times to be customer specific and pre-generate all customers along with their associated service times, their probability of requiring technical support, and their patience time. The same set of pre-generated customers are used to simulate a given day (replication) under different staffing configurations. 

\begin{figure}[!htb]
    \begin{subfigure}[b]{0.48\textwidth}
        \includegraphics[width=\textwidth]{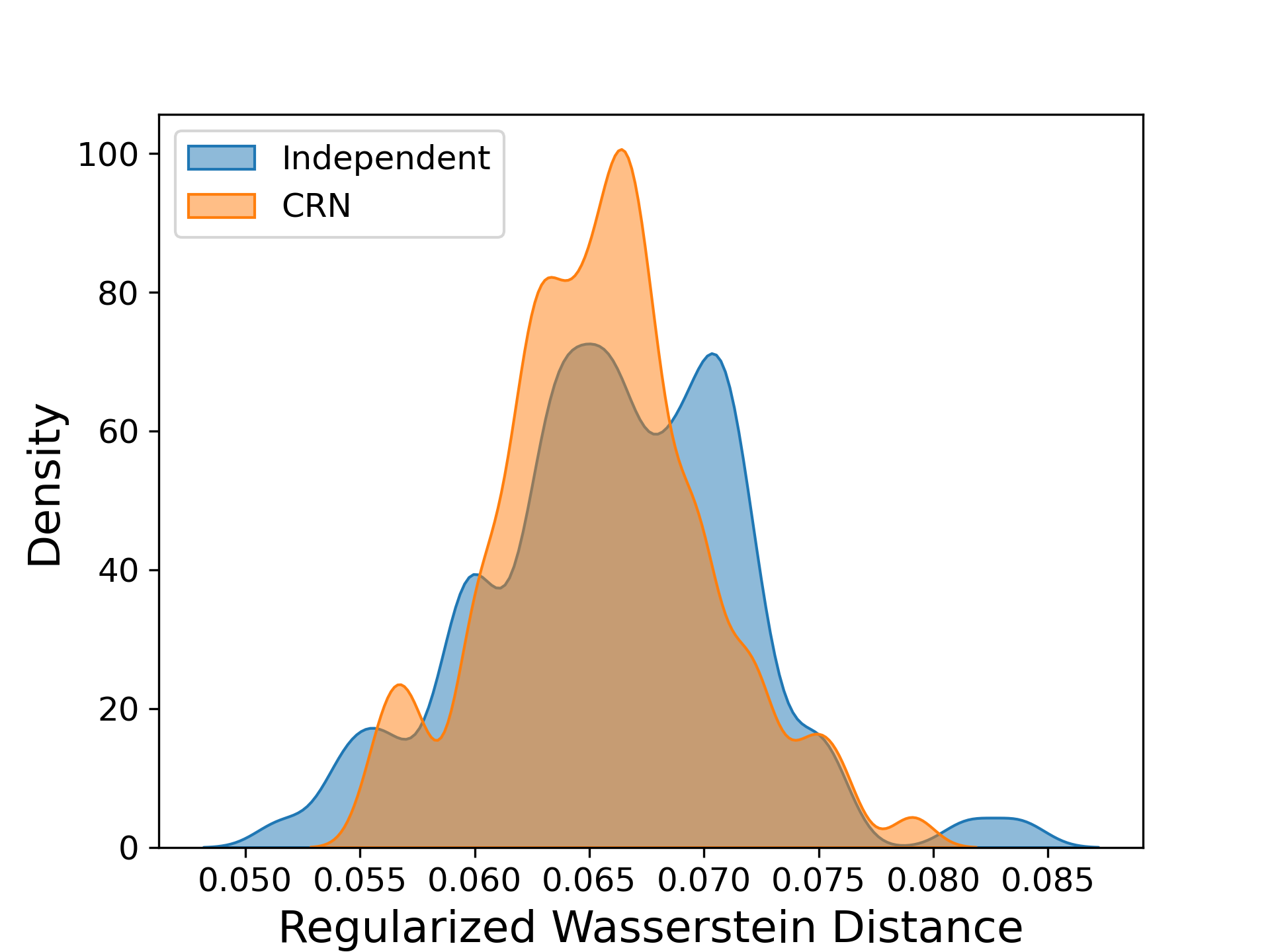}
        \label{fig: CRN distance}
    \end{subfigure}
    \begin{subfigure}[b]{0.48\textwidth}
        \includegraphics[width=\textwidth]{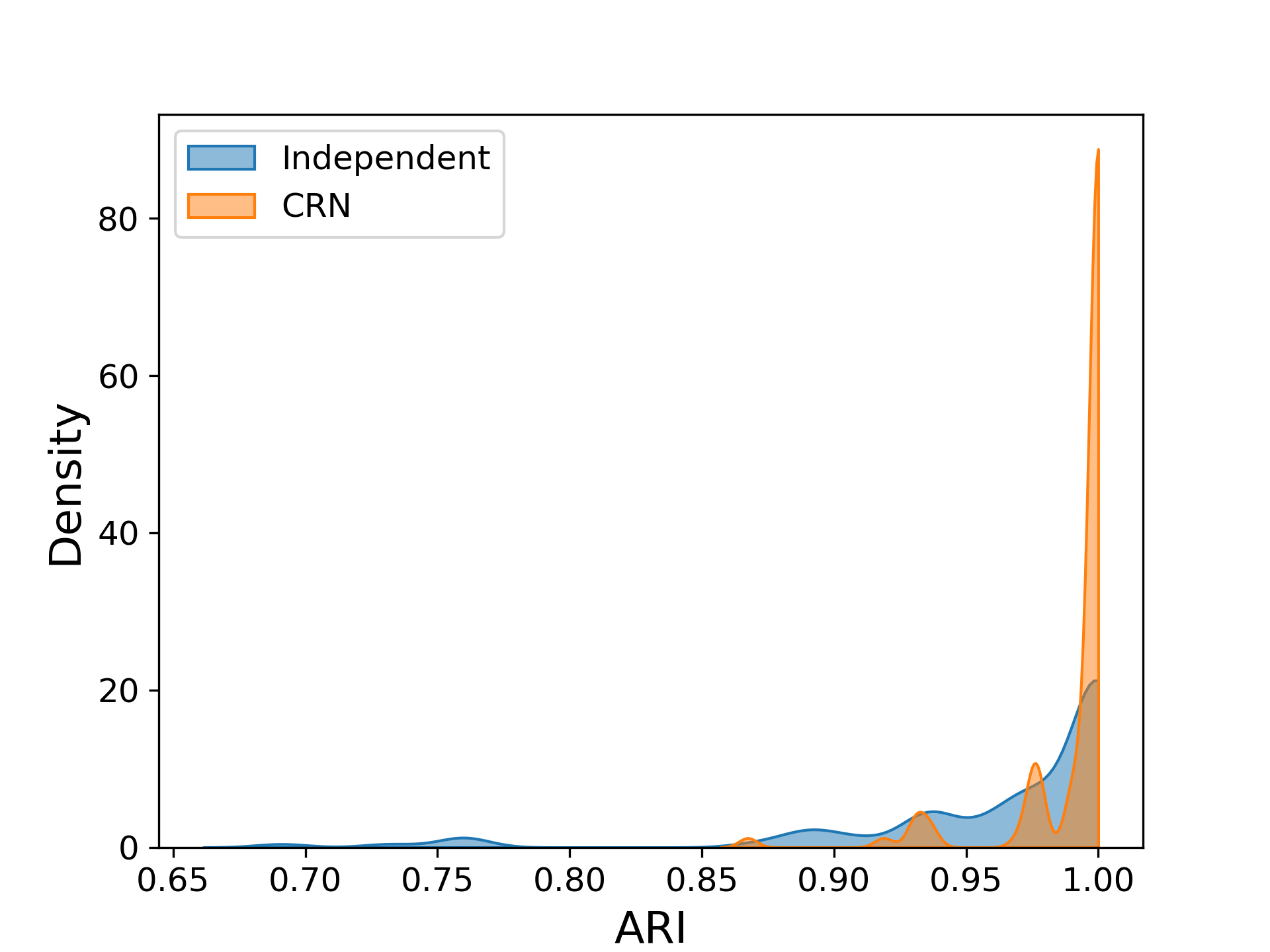}
        \label{fig: ARI}
    \end{subfigure}
    \caption{(Left) distribution of regularized Wasserstein distance between the empirical distributions for staffing configurations  $(9, 23, 27)$ and $(7, 28, 14)$ based on 100 macroreplications for CRN and independent sampling. (Right) Adjusted Rand Index (ARI) for clusterings resulting from CRN and independent sampling.}
    \label{fig: CRN figures}
\end{figure}

We first test whether CRN can reduce the variability of the regularized Wasserstein distance.
We simulate 40 days under two staffing configurations, $(9, 23, 27)$ and $(7, 28, 14)$---once independently and once using CRN---and in each case calculate the regularized Wasserstein distance between the two empirical distributions. This process is repeated 100 times and the distributions of the regularized Wasserstein distance under both settings are plotted in Figure \ref{fig: CRN figures}. 
The estimated variances of the regularized Wasserstein distance for the independent sampling and CRN cases are $3.26 \times 10^{-5}$ and $2.15 \times 10^{-5}$, respectively, and an $F$-test testing whether the two variances are equal yields a $p$-value of $0.0198$.

We next study whether CRN can affect how the clusterings produced by our algorithm in Section 4.1 compare with the \emph{true} clustering of the scenarios' output distributions, were those distributions known.
To obtain a proxy for the true clustering, we simulate each staffing configuration 2000 times, apply our agglomerative clustering algorithm, and treat the result as the true clustering. For both the independent and CRN cases, we simulate each staffing configuration over 15 days, and conduct 100 macroreplications.
To measure the similarity between two clusterings, we employ the Adjusted Rand Index (ARI) introduced by \cite{hubert1985comparing}.
The ARI ranges from $-1$ to $1$, where $1$ indicates the clusterings match perfectly.
The distributions of the ARI using independent sampling and CRN are depicted in Figure \ref{fig: CRN figures}. The mean ARI for clusterings with independent sampling is approximately 0.9461 with a standard deviation of 0.0736, whereas the mean ARI for clusterings using CRN is approximately 0.9845, with a lower standard deviation of 0.0194.
These results provide some empirical evidence that, compared to independent sampling, using CRN can produce clusterings that are not only more consistent with the ground truth but also less variable.

\section{Conclusion}
\label{sec:conclusion}
This paper introduces an efficient agglomerative clustering algorithm for multivariate empirical distributions, motivated by the setting of analyzing simulation output data. Clustering simulation output data by scenario can be a powerful approach for anomaly detection, pre-optimization, and online monitoring. The proposed algorithm demonstrates superior computational efficiency in terms of both the choice of distance metric (regularized Wasserstein) and the clustering methodology (agglomerative). Limited experiments suggested that synchronizing random numbers across scenarios, \`{a} la CRN, can both reduce the variability of the returned clustering and improve its accuracy. Also A Python package for the agglomerative clustering algorithm is under development. Future research directions include clustering simulation output distributions in a streaming-data setting and clustering the curves of state variables over time for discrete-event simulations, which can provide deeper insights into dynamic system behavior. 


\bibliographystyle{named} 
\bibliography{refs}

\end{document}